\documentclass{article}
\usepackage{spconf,amsmath,graphicx,amsfonts,multirow}

\title{Fast Cross-Correlation for TDoA Estimation on Small Aperture Microphone Arrays}
\name{Fran\c{c}ois Grondin, Marc-Antoine Maheux, Jean-Samuel Lauzon, Jonathan Vincent, Fran\c{c}ois Michaud}
\address{Université de Sherbrooke (Québec, Canada)}
\begin{document}
%
\maketitle
\begin{abstract}
This paper introduces the Fast Cross-Correlation (FCC) method for Time Difference of Arrival (TDoA) Estimation for pairs of microphones on a small aperture microphone array.
FCC relies on low-rank decomposition and exploits symmetry in even and odd bases to speed up computation while preserving TDoA accuracy.
FCC reduces the number of flops by a factor of 4.5 and the execution speed by factors between 3.5 and 8.3 on embedded hardware, compared to the state-of-the-art Generalized Cross-Correlation (GCC) method that relies on the Fast Fourier Transform (FFT).
This improvement can provide portable microphone arrays with extended battery life and allow real-time processing on low-cost hardware.
\end{abstract}
\section{Introduction}
\label{sec:intro}

Many smart devices (e.g., smart speakers, tablets) are equipped with multiple microphones to perform sound source localization and speech enhancement \cite{barker2015third}.
This comes particularly handy to cope with performance drop for distant speech recognition tasks \cite{tang2018study}.
While many devices perform recognition tasks on the cloud, there is a growing need for local processing, to reduce bandwidth usage, minimize latency and protect privacy.
This requires to revisit some algorithms to reduce the computational load and extend battery life.

Sound source localization (SSL) consists in estimating the direction of arrival (DoA) of sound using multiple microphones positioned in a specific geometrical configuration.
There are three main categories of SSL methods: 1) beamforming, 2) subspace decomposition, and 3) machine learning approaches.
Beamforming consists in pointing a beam at potential DoAs around the microphone array to find the direction that captures the most power.
This can be achieved using a Steered-Response Power Phase Transform beamformer, referred to SRP-PHAT \cite{do2007real,cobos2010modified}.
It is common to compute SRP-PHAT with the TDoA obtained using the Generalized Cross-Correlation with Phase Transform (GCC-PHAT) for each pair of microphones \cite{valin2007robust}.
However, scanning every possible DoA involves numerous lookups, especially when performing 3-D localization.
SMP-PHAT method exploits the symmetry of the microphone array to reduce the number of computations \cite{grondin2022smp}.
It is also possible to use the SVD-PHAT method, which relies on Singular Value Decomposition of the SRP-PHAT projection matrix \cite{grondin2019svd}.
SVD-PHAT however requires to compute offline a decomposition for each specific microphone geometry, which can be computationally expensive for arrays with numerous microphones, or if the shape changes dynamically over time.

Subspace decomposition methods include Multiple Signal Classification (MUSIC) \cite{schmidt1986multiple}.
Formerly used for narrowband signals, MUSIC was adapted to broadband signals like speech, and can be defined as Standard Eigenvalue Decomposition MUSIC (SEVD-MUSIC) \cite{ishi2009evaluation, bando2021robust}.
SEVD-MUSIC however assumes the speech signal is more powerful than noise at each frequency bin.
Nakamura et al. introduced the MUSIC based on Generalized Eigenvalue Decomposition (GEVD-MUSIC) method to overcome this limitation \cite{nakamura2009intelligent, nakadai2012robot}.
While this method handles noisy scenarios, it also introduces some localization errors because the generated noise subspace is made of correlated bases.
To deal with this issue, Generalized Singular Value Decomposition (GSVD-MUSIC) enforces orthogonality between the noise subspace bases and thus improves the DoA estimation accuracy \cite{nakamura2012real}. 
All MUSIC-based methods rely on computationally expensive eigenvalue or singular value decompositions, which make on-board processing challenging.

Machine learning approaches usually involve training a deep neural network with a specific array geometry and predict a class amongst all classes that stand for all potential DoAs.
Chakrabarty and Habets \cite{chakrabarty2017broadband, chakrabarty2019multi} demonstrate that a convolutional neural network (CNN) could be trained on white noise signals to estimate the DoA for a Uniform Linear Array (ULA).
Other approaches also demonstrate the robustness of CNNs and fully connected networks in low signal-to-noise ratio (SNR) scenarios \cite{papageorgiou2021deep,ozanich2020feedforward}.
\c{C}ak{\i}r et al. \cite{cakir2017convolutional} showed that convolutive and recursive neural networks (CRNNs) could be used for sound event detection, followed by Adavanne et al. \cite{adavanne2018sound} who demonstrated CRNNs could estimate the DoAs for a specific class of sound.
It is also possible to estimate the TDoAs for each pair of microphones for a specific class of sound using CRNNs \cite{grondin2019sound}.
While most deep-learning-based approaches outperform traditional signal processing methods in terms of accuracy, they involve a significant amount of computations due to the numerous parameters in their layers, which makes them less suitable for real-time applications on low-cost embedded hardware.

This paper presents a novel method to estimate the TDoA (which can be the building block for DoA estimation), called Fast Cross-Correlation (FCC), specifically adapted to microphone arrays with aperture of few centimeters with limited available computing capacity.
The two main contributions of this paper are: 1) it describes the FCC method and demonstrates it reduces the theoretical amount of flops compared to the GCC approach; 2) it explains FCC implementation in the C language and demonstrates its superior performance in term of execution time on low-cost hardware.

\section{Time Difference of Arrival}
\label{sec:tdoa}

The TDoA corresponds to the time delay due to the propagation of sound in air between two microphones (in seconds or in samples).
For a given direction of arrival (DoA) of sounds denoted by the angle $\theta \in [0,\pi]$ in radians, the TDoA (in samples) corresponds to $\tau = f_S \left( d/c \right) \cos \theta$, where $c \in \mathbb{R}^+$ stands for the speed of sound (in m/sec), $d \in \mathbb{R}^+$ for the distance between both microphones (in m), $f_S \in \mathbb{N}$ for the sample rate (in samples/sec) and $\tau \in [-\tau_{max},+\tau_{max}]$ for the TDoA, where $\tau_{max} = d f_S / c$.

\begin{figure*}[!ht]
    \centering
    \includegraphics[width=0.9\linewidth]{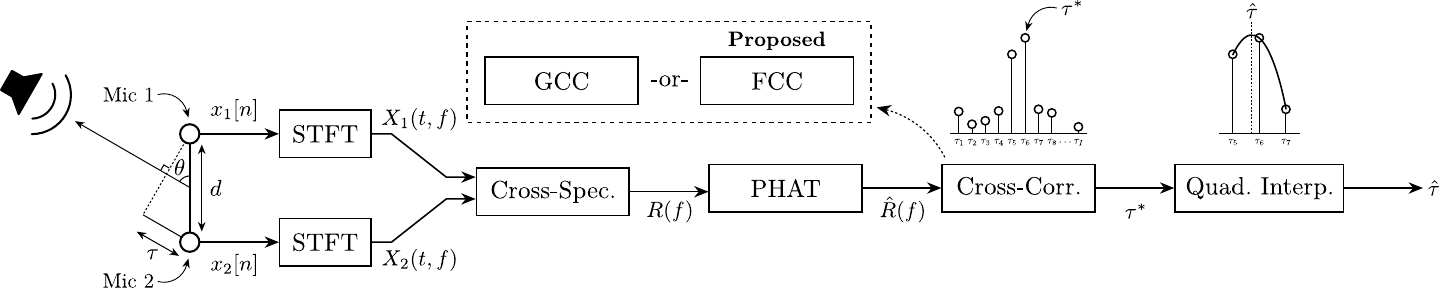}
    \caption{TDoA estimation, using either the traditional method (GCC) or the proposed approach (FCC).}
    \label{fig:tdoa_overview}
    \vspace{-10pt}
\end{figure*}

Figure \ref{fig:tdoa_overview} shows the overall processing pipeline to estimate the TDoA from two microphones.
We define $x_1[n] \in \mathbb{R}$ and $x_2[n] \in \mathbb{R}$ as the input signals in the time-domain, where $n \in \mathbb{N}$ stands for the sample index.
A Short-Time Fourier Transform (STFT) with a frame size of $N \in \{2, 4, 8, 16, \dots\}$ samples is performed on each signal.
Here the STFT uses a Hann window and frames overlap by $50\%$.
This generates the frames $X_1(t,f) \in \mathbb{C}$ and $X_2(t,f) \in \mathbb{C}$, where $t \in \{1, 2, \dots, T\}$ is the frame index, $T \in \mathbb{N}$ the number of frames, and $f \in \{0, 1, \dots, N/2\}$ the frequency bin index.
The cross-spectrum $R(t,f) \in \mathbb{C}$ is estimated recursively over time, using a smoothing factor defined by the parameter $\alpha \in [0,1]$:
\begin{equation}
    R(t,f) = (1-\alpha) R(t-1,f) + \alpha X_1(t,f)X_2(t,f)^*,
    \label{eq:xspec}
\end{equation}
where $\{\dots\}^*$ stands the complex conjugate.
The phase transform normalizes the cross-spectrum, such that $\hat{R}(t,f) = R(t,f)/|R(t,f)|$.
From now on we can omit the index $t$ for clarity (so $\hat{R}(t,f)$ becomes $\hat{R}(f)$), without loss of generality.

The expression $y(\tau)$ stands for the cross-correlation (computed on $\hat{R}(f)$, either with GCC or FCC) for each delay $\tau \in \{\tau_1, \tau_2, \dots, \tau_I\}$, where $I$ stands for the number of candidates.
The goal is to find the delay $\tau^*$ that maximizes $y(\tau)$.
Once $\tau^*$ found, $\hat\tau$ is calculated using a polynomial interpolation to improve accuracy.
There are various interpolation methods, but here we limit ourselves to quadratic interpolation, as it requires little computation and has a closed-form solution \cite{abe2004design}. 

\section{Generalized Cross-Correlation}
\label{sec:gcc}

The Generalized Cross-Correlation is widely used for TDoA estimation and is appealing as it relies on the well-known Fast Fourier Transform (FFT) algorithm.
The cross-correlation is computed as follows, where $\Re\{\dots\}$ captures the real part:
\begin{equation}
    y(\tau) = 2\Re\left\{\sum_{f=0}^{N/2}{\hat{R}(f)\exp(j 2 \pi f \tau/N)}\right\}.
    \label{eq:y}
\end{equation}

A real-valued Inverse Fast Fourier Transform (IFFT) can be used to compute efficiently the expression in (\ref{eq:y}).
However, $\tau$ needs to be an integer, which reduces considerably the number of TDoA candidates, especially when the distance $d$ between microphones is small.
This also becomes an issue for quadratic interpolation, as there is only one discrete delay value that samples the main lobe.
The solution to this limitation consists in performing an IFFT twice the size of the input frame.
This is achieved by padding the frame with zeros to double the number of frequency bins ($r=2$) prior to computing the IFFT.
This results in time-domain interpolation, with a resolution of 0.5 sample (or more generally, $1/r$ sample).
This operation allows sampling the main lobe properly for accurate estimation of the TDoA with quadratic interpolation.
It also implies there are $I = 4\lceil\tau_{max}\rceil + 1$ delay candidates, where $\lceil\dots\rceil$ stands for the ceiling operator.
For example, to deal with distance up to $d=0.15$ m with $f_S = 16000$ samples/sec and $c=343.0$ m/sec, we get $\tau_{max}=8$ and $I=33$.

Interpolation can become an issue for real-time systems as the computational load increases when the IFFT size gets larger.
Although the exact number of flops depends on the specific FFT algorithm implementation, it is in the order of $5N/r \log_2 (rN)$ for radix-2 implementation with real numbers and $rN$ samples \cite{johnson2006modified}.
With a frame size of $N=512$ and $r=2$, GCC requires approximately $25600$ flops.

\section{Fast Cross-Correlation}
\label{sec:fcc}

\begin{figure}[!ht]
    \centering
    \includegraphics[width=0.8\linewidth]{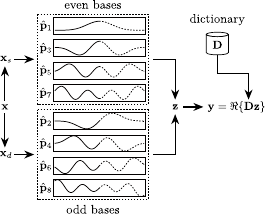}
    \caption{Overview of the FCC-PHAT approach. Bases and dictionary are computed offline and stored in memory.}
    \label{fig:fcc}
    \vspace{-15pt}
\end{figure}

Figure \ref{fig:fcc} shows the pipeline of the proposed FCC approach.
Equation \ref{eq:y} is formulated in vector form in this section to ease the description of FCC.
All the cross-spectrum coefficients with phase transform can be stacked for all frequency bins ($f \in \{0, 1, \dots, N/2\}$) in a single vector called $\mathbf{x} \in \mathbb{C}^{N/2+1}$.
Similarly, all the beamformer coefficients $w(\tau,f) = \exp(j 2 \pi f \tau/N)$ are concatenated in a vector $\mathbf{w}(\tau) \in \mathbb{C}^{N/2+1}$.
The vectors can be combined in a matrix $\mathbf{W} \in \mathbb{C}^{I \times N/2+1}$:
\begin{equation}
    \mathbf{W} = \left[ 
    \begin{array}{cccc}
        \mathbf{w}(\tau_1) &
        \mathbf{w}(\tau_2) &
        \dots &
        \mathbf{w}(\tau_I) \\
    \end{array}
    \right]^T.
    \label{eq:Wconcat}
\end{equation}

The expression in (\ref{eq:y}) can be represented in matrix form as $\mathbf{y} = 2\Re \{ \mathbf{W} \mathbf{x} \}$ where $\mathbf{y} \in \mathbb{C}^{I}$ corresponds to the cross-correlation $y(\tau)$ for each delay $\tau \in \{\tau_1, \tau_2, \dots, \tau_I\}$:
\begin{equation}
    \mathbf{y} = \left[ 
    \begin{array}{cccc}
        y(\tau_1) &
        y(\tau_2) &
        \dots &
        y(\tau_I) \\
    \end{array}
    \right]^T.
\end{equation}

This involves $I(N/2+1)$ complex multiplications (CMs), and $IN/2$ complex additions (CAs).
Each CM involves $4$ real multiplications (RMs), and $2$ real additions (RAs), while each CA requires $2$ RAs.
This sums up to $I(4N+6)$ flops, which can be prohibitive for real-time applications when the number of candidates $I$ gets large.
With $\tau_{max} = 8$ and frame size of $N=512$ samples, the direct computation involves $67782$ flops.
This is more flops than with GCC, which suggests further optimization is required.
As explored previously in \cite{grondin2019svd}, the matrix $\mathbf{W}$ is usually low-rank, especially when the distance $d$ between the microphones is in the order of a few centimeters.
This matrix is factored using singular value decomposition (SVD), such as $\mathbf{W} = \mathbf{U} \mathbf{S} \mathbf{V}^H$, where $\mathbf{U} \in \mathbb{C}^{I \times I}$, $\mathbf{S} \in \mathbb{R}^{I \times N/2+1}$, $\mathbf{V} \in \mathbb{C}^{N/2+1 \times N/2+1}$ and $\{\dots\}^H$ stands for the Hermitian transpose operator.

As shown in \cite{grondin2019svd}, the matrices can be cropped by keeping only the $K$ most significant singular values.
This leads to matrices $\hat{\mathbf{U}} \in \mathbb{C}^{I \times K}$, $\hat{\mathbf{S}} \in \mathbb{R}^{K \times K}$ and $\hat{\mathbf{V}} \in \mathbb{C}^{N/2+1 \times K}$, such that $\mathbf{W} \approx \hat{\mathbf{U}}\hat{\mathbf{S}}\hat{\mathbf{V}}^H$.
The parameter $K$ is chosen to ensure accurate reconstruction of $\mathbf{W}$, to be set experimentally.
For now, we assume $K$ is known and we define the projection matrix $\mathbf{P} \in \mathbb{C}^{K \times N/2+1}$ and dictionary $\mathbf{D} \in \mathbb{C}^{I \times K}$ as $\mathbf{P} = \hat{\mathbf{S}} \hat{\mathbf{V}}^H$ and $\mathbf{D} = \hat{\mathbf{U}}$.
The dictionary is the concatenation of rows $\mathbf{d}(i) \in \mathbb{C}^{1\times K}$ for each TDoA $\tau_i$.
The cross-correlation corresponds to $\mathbf{y} = 2\Re\{\mathbf{D}\mathbf{z}\}$, where $\mathbf{z}=\mathbf{Px}$.
At this point, computing $\mathbf{z}$ involves $K(N/2+1)$ CMs, and $KN/2$ CAs, for a total of $K(4N+6)$ flops.
The computation of $2\Re\{\mathbf{D}\mathbf{z}\}$ then involves $2IK$ RMs, and $I(2K-1)$ RAs, for a total of $4KI-I$ flops.
This sums up to $K(4N+6) + I(4K-1)$ flops, which leads approximately to a computational load reduction by a factor of $N/K$ when $I \ll N$.
With the previous example ($\tau_{max}=8$ so $I=33$, and $N=512$), and by choosing $K=8$, the number of flops goes down to $17455$ (compared to $67782$). 
We can reduce further the number of flops by exploiting some properties of the bases generated by SVD.
The projection matrix $\mathbf{P}$ is expressed as $K$ concatenated vectors:
\begin{equation}
    \begin{array}{c}
        \mathbf{P} = \left[ 
        \begin{array}{cccc}
            \mathbf{p}_1 &
            \mathbf{p}_2 &
            \dots &
            \mathbf{p}_K \\
        \end{array}
        \right]^T, \\
        \vspace{-7pt}\\
        \mathbf{p}_k = \left[
        \begin{array}{cccc}
        p(k,0) & p(k,1) & \dots & p(k,N/2)
        \end{array}
        \right].
    \end{array}
\end{equation}

All the bases are either even and real, or odd and purely imaginary, due to the even and odd vectors in (\ref{eq:Wconcat}), such that:
\begin{equation}
    p(k,f) = \begin{cases}
        +p(k,N/2-f) & k = \{1,3,5,\dots \} \\
        -p(k,N/2-f) & k = \{2,4,6,\dots \} \\
    \end{cases},
\end{equation}
where $p(k,f) \in \mathbb{R}$ when $k \in \{1, 3, 5, \dots\}$, and $p(k,f) \in j\mathbb{R}$ when $k \in \{2, 4, 6, \dots\}$.
The second half is therefore redundant and the vectors can be cropped to $\hat{\mathbf{p}}_k \in \mathbb{C}^{N/4+1}$:
\begin{equation}
    \hat{\mathbf{p}}_k = \left[
    \begin{array}{cccc}
        p(k,0) & p(k,1) & \dots & p(k,N/4)
    \end{array}
    \right].
\end{equation}

Frame $\mathbf{x} \in \mathbb{C}^{N/2+1}$ can be reorganized as the sum ($\mathbf{x}_{s}$) and difference ($\mathbf{x}_{d}$) of the cross-spectrum coefficients:
\begin{equation}
    \{\mathbf{x}_{s},\mathbf{x}_{d}\} = \left[
    \begin{array}{c}
        \hat{R}(0) \pm \hat{R}(N/2) \\
        \vdots \\
        \hat{R}(N/4-1) \pm \hat{R}(N/4+1) \\
        \hat{R}(N/4) \\
    \end{array}
    \right].
\end{equation}

\begin{table*}[!ht]
    \centering
    \caption{Mean absolute error (MAE) in degrees ($^{\circ}$)}
    \renewcommand{\arraystretch}{1.05}
    \begin{tabular}{|c|c|cc|ccccccccccc|}
        \hline
        \multirow{2}{*}{Method} & \multirow{2}{*}{Flops} & \multicolumn{2}{c|}{Run time ($\mu$sec)} & \multicolumn{11}{c|}{Distance between microphones ($d$) (in m)}\\
         & & RPZ & RP4 & 0.05 & 0.06 & 0.07 & 0.08 & 0.09 & 0.10 & 0.11 & 0.12 & 0.13 & 0.14 & 0.15 \\
        \hline
        \hline
         GCC ($r=1$) & 11520 & 50.3 & 3.9 & 9.4 & 8.7 & 8.1 & 7.8 & 7.8 & 7.2 & 7.5 & 8.0 & 8.3 & 8.7 & 8.5 \\
         GCC ($r=2$) & 25600 & 206.6 & 7.5 & 8.0 & \textbf{7.6} & 7.4 & \textbf{7.1} & \textbf{7.1} & \textbf{6.5} & \textbf{6.6} & \textbf{7.3} & \textbf{7.8} & \textbf{7.8} & \textbf{7.7} \\
         GCC ($r=4$) & 56320 & 455.9 & 15.9 & \textbf{7.9} & 7.7 & \textbf{7.2} & 7.3 & 7.2 & 6.7 & 6.8 & 7.5 & \textbf{7.8} & 8.0 & \textbf{7.7} \\
        \hline
         FCC ($K=1$) & 1125 & 8.6 & 1.0 & 20.3 & 26.5 & 28.6 & 29.4 & 29.7 & 29.2 & 28.1 & 27.5 & 26.9 & 29.5 & 29.4 \\
         FCC ($K=2$) & 1771 & 9.7 & 1.2 & 10.1 & 10.4 & 11.9 & 15.6 & 17.9 & 18.7 & 19.7 & 20.5 & 20.8 & 23.1 & 24.2 \\
         FCC ($K=3$) & 2417 & 12.1 & 1.4 & 8.6 & 8.8 & 9.4 & 10.0 & 10.9 & 12.5 & 14.2 & 14.7 & 15.9 & 18.0 & 19.6 \\
         FCC ($K=4$) & 3063 & 15.2 & 1.5 & 9.2 & 8.7 & 7.9 & 8.3 & 8.9 & 8.9 & 9.7 & 10.7 & 12.8 & 14.7 & 16.7 \\
         FCC ($K=5$) & 3709 & 17.9 & 1.7 & 8.2 & 8.1 & 8.0 & 7.7 & 7.2 & 7.1 & 8.0 & 
         8.7 & 9.4 & 10.7 & 12.6 \\
         FCC ($K=6$) & 4355 & 19.2 & 1.9 & 8.4 & 7.8 & 7.4 & 7.5 & 7.6 & 6.9 & 7.1 & 7.6 & 8.4 & 8.9 & 9.0 \\
         FCC ($K=7$) & 5001 & 21.7 & 2.0 & 8.0 & 8.0 & 7.6 & \textbf{7.1} & \textbf{7.0} & 6.6 & 6.9 & 7.5 & \textbf{7.7} & 8.1 & 7.9 \\
         FCC ($K=8$) & 5647 & 25.0 & 2.1 & \textbf{7.9} & \textbf{7.6} & \textbf{7.3} & 7.3 & 7.2 & \textbf{6.5} & \textbf{6.7} & \textbf{7.4} & 7.9 & \textbf{7.8} & \textbf{7.4} \\
        \hline
    \end{tabular}
    \label{tab:mae}
\end{table*}

Computing $\mathbf{x}_{s}$ and $\mathbf{x}_{d}$ only needs to be performed once for a given $\mathbf{x}$, and can be stored in cache memory and reused multiple times.
This new formulation implies that $\mathbf{z}$ can be computed as follows, where $\cdot$ is the dot product:
\begin{equation}
    \mathbf{z} = \left[
    \begin{array}{cccc}
    \hat{\mathbf{p}}_1 &
    \hat{\mathbf{p}}_2 &
    \hat{\mathbf{p}}_3 &
    \dots \\
    \end{array}
    \right] \cdot
    \left[
    \begin{array}{cccc}
    \mathbf{x}_{s} & \mathbf{x}_{d} & \mathbf{x}_{s} & \dots \\
    \end{array}
    \right].
    \label{eq:Px_odd_even}
\end{equation}

The bases now hold $N/4+1$ elements (instead of $N/2+1$) and all the elements of $\hat{\mathbf{p}}_k$ are purely real or imaginary numbers.
Computing (\ref{eq:Px_odd_even}) involves $K(N/2+2)$ RMs and $KN/2$ RAs, for a total of $K(N+2)$ flops.
Computing the vectors $\mathbf{x}_{s}$ and $\mathbf{x}_{d}$ also adds $N$ flops, which leads to a total of $K(N+2)+N$ flops.
Computing $\mathbf{D}\mathbf{z}$ adds $I(4K-1)$ flops, and leads to $K(N+2)+N+I(4K-1)$ flops in total.
With the previous example ($\tau_{max}=8$ so $I=33$, and $N=512$), the number of flops drops to $5647$, which represents a significant gain by a factor of $4.5$ compared to GCC ($25600$ flops).

\section{Results}
\label{sec:results}

The BIRD dataset provides simulations performed in virtual rooms with width and length between $5$ and $15$ m, and height between $3$ and $4$ m \cite{grondin2020bird}.
Room Impulse Responses (RIRs) that correspond to an absorption coefficient of the surface between $0.2$ and $0.8$ are used, which leads to reverberation times RT60 between $0.10$ and $1.00$ sec that follow a gamma distribution \cite{grondin2020bird}.
The pair of microphones is positioned randomly in the virtual room, with a spacing between microphones that correspond to $d \pm 0.001$ m, where $d$ lies in the range $[0.01,0.15]$ m.
The sound source is positioned randomly, and kept at a distance between $1$ and $5$ m from the microphones.
The speed of sound varies between $335$ and $350$ m/sec, to accomodate for different ambiant temperatures.
White noise is convolved with the RIRs to generate the signals at both microphones.

The accuracy and computational load of FCC depends on the parameter $K \in \{1, 2, \dots, 8\}$.
We set $\tau_{max} = 8$ to generate a set of bases $\mathbf{P}$ suitable for spacing up to $d=0.15$ m at a sample rate of $f_S = 16000$ samples/sec and a minimum speed of sound of $c = 335.0$ m/sec.
GCC provides a baseline, with different time interpolation for $r \in \{1, 2, 4\}$.
Quadratic interpolation is then performed to refine the TDoA estimation.
The frame size is $N=512$ samples and the smoothing factor is $\alpha=0.1$.
The mean absolute error (MAE) in degrees ($^{\circ}$) is:
\begin{equation}
    \mathrm{MAE} = \frac{1}{T}\sum_{t=1}^{T}{|\theta_t - \hat{\theta}_t|},
\end{equation}
and where $\theta = \arccos{\left(\tau_t c/(d f_S)\right)}$.
Table \ref{tab:mae} presents the mean absolute error (MAE) for GCC (with different values of $r$) and FCC (with different values of $K$), with RT60 values selected randomly at each simulation.
The results show the need to increase the IFFT size ($r=2$) with GCC to get a good accuracy.
The difference between $r=2$ and $r=4$ is negligible, suggesting a larger $r$ increases the computational load without improving the accuracy significantly.
Note that FCC with $K=8$ can achieve a similar accuracy as GCC (with $r=2$) for all configurations.

While the number of flops provides a rough estimation of the computational load, it ignores other factors such as the memory access time and the potential vector optimizations when the algorithm runs on a modern processor.
To consider these factors, FCC is implemented in the C language\footnote{https://github.com/francoisgrondin/fcc} and runs on a Raspberry Pi Zero (RPZ) and a Raspberry Pi 4 (RP4) devices.
GCC runs on the same hardware, and uses the optimized FFTW library to compute the IFFT efficiently \cite{johnson2006modified}.
Both methods are compiled with the GNU Compiler Collection and use vectorization.
Table \ref{tab:mae} shows that FCC ($K=8$) runs 8.3 and 3.5 times faster than GCC ($r=2$) for the RPZ and RP4, respectively.
This confirms that FCC runs faster when deployed on low-cost hardware.

\section{Conclusion}
\label{sec:conclusion}

This paper introduces FCC, well-suited for microphone arrays with an aperture of a few centimeters.
FCC provides a similar accuracy to GCC but outperforms it in terms of computational load.
We demonstrate that the execution time goes down by factors of 8.3 and 3.5 on RPZ and RP4, respectively.
The next step consists in using FCC to estimate TDoAs for each pair of microphones in a microphone array, and to perform a search for the best DoA candidate.
This functionality could be added to existing frameworks such as ODAS \cite{grondin2021odas}, and exploits optimization strategies introduced with SMP-PHAT \cite{grondin2022smp}.

\bibliographystyle{IEEEbib}
\bibliography{biblio}

\end{document}